\documentclass[sigconf]{acmart}

\usepackage{booktabs}
\usepackage{todonotes}
\usepackage[font=small,skip=3pt]{caption}

\copyrightyear{2022}
\acmYear{2022}
\setcopyright{rightsretained}
\acmConference[SAC '22]{The 37th ACM/SIGAPP Symposium on Applied Computing}{April 25--29, 2022}{Virtual Event, }
\acmBooktitle{The 37th ACM/SIGAPP Symposium on Applied Computing (SAC '22), April 25--29, 2022, Virtual Event, }\acmDOI{10.1145/3477314.3506963} \acmISBN{978-1-4503-8713-2/22/04}

\acmArticle{4}
\acmPrice{15.00}
\begin{document}
\title{Student Research Abstract: Microservices-based Systems Visualization}
\author{Amr S. Abdelfattah}
\orcid{1234-5678-9012}
\affiliation{%
  \institution{Baylor University}
  \streetaddress{P.O. Box 1212}
  \city{Waco} 
  \state{Texas} 
  \postcode{43017-6221}
}
\email{amr_elsayed1@baylor.edu}
\begin{abstract}
The evolution of decentralized microservice-based systems is challenging. These challenges are classified into static and dynamic categories. Regarding the static perspective, documenting and visualizing the fluid application topology is something few have been able to accomplish. Building an architecture map of services design is a complicated task in its interpretation rather than construction. Therefore, the system-centric and up-to-date view became essential for such distributed systems.
The dynamic perspective considers the process of investigation and service path detection. Therefore performing root cause analysis is a burdening task; such that tracing data is needed to be put in the right context to facilitate the investigation. Moreover, visualizing these traces over the traditional visualization techniques couldn't be feasible with the large number of microservices involved in the system.
This paper proposes a visualization concept for microservices-based systems using the Augmented Reality (AR) technique, which merges these static and dynamic behaviors into a single centric view. In addition, we challenge the difficulty related to tracing and debugging an issue in such distributed systems. This concept is designed to work as a dynamic documentation and traceability platform for these systems. A proof of concept and a research study are implemented to show the viability and success of this proposal.
\end{abstract}
\maketitle
\section*{PROBLEM AND MOTIVATION}
Microservice solutions are widely used but the fundamental tools and perspectives to better observe and understand these systems are still missing.
The issues investigation process involves a sequence of actions as follows: getting back to the services documentation —which may be outdated— that should describe the accused microservice and its dependencies, figuring out the system logs and traces data, trying to guess the failed scenario, then simulating it to ensure the root cause of such issue.
That requires engineers to understand what happened across the entire service graph at the time of the debugging, which seems to counter to the ethos of microservices architectures in the first place.\\
Bogner et al. focused on this challenge in a systematic grey literature review \cite{bogner2021industry}. They concluded that most painful challenges in the microservices industry are the Service-cutting and No-System-Centric views. Furthermore, these challenges are vital, especially for large systems, for example, Amazon.com calls between 100-150 web services to build a page. As well as Netflix microservices architecture supports 5 billion services calls per day, 99.7\% of which are internal \cite{appdynamics2015}.
Therefore, what is ideally required at the time of debugging is a system-centric view that will help reduce the search space and provide living documentation so that engineers can reach the root cause with less effort.\\
One way to address this is to use the system itself as living documentation and, consequently, visualize it in a tailored and detailed way to show all the required information.
The visual space limitation especially in the 2D modeling is a critical challenge for visualizing these architectures. Therefore, a novel approach is needed to face these challenges in a tailored way for microservices nature. The proposed approach involves Augmented Reality (AR) in this context, which has multiple benefits, either for providing a limitless rendering space, or for supporting collaborative activities which are essential for system analysis practices.
\begin{figure*}[!htb]
    \centering
    \begin{minipage}{.23\textwidth}
        \centering
        \includegraphics[width=\linewidth]{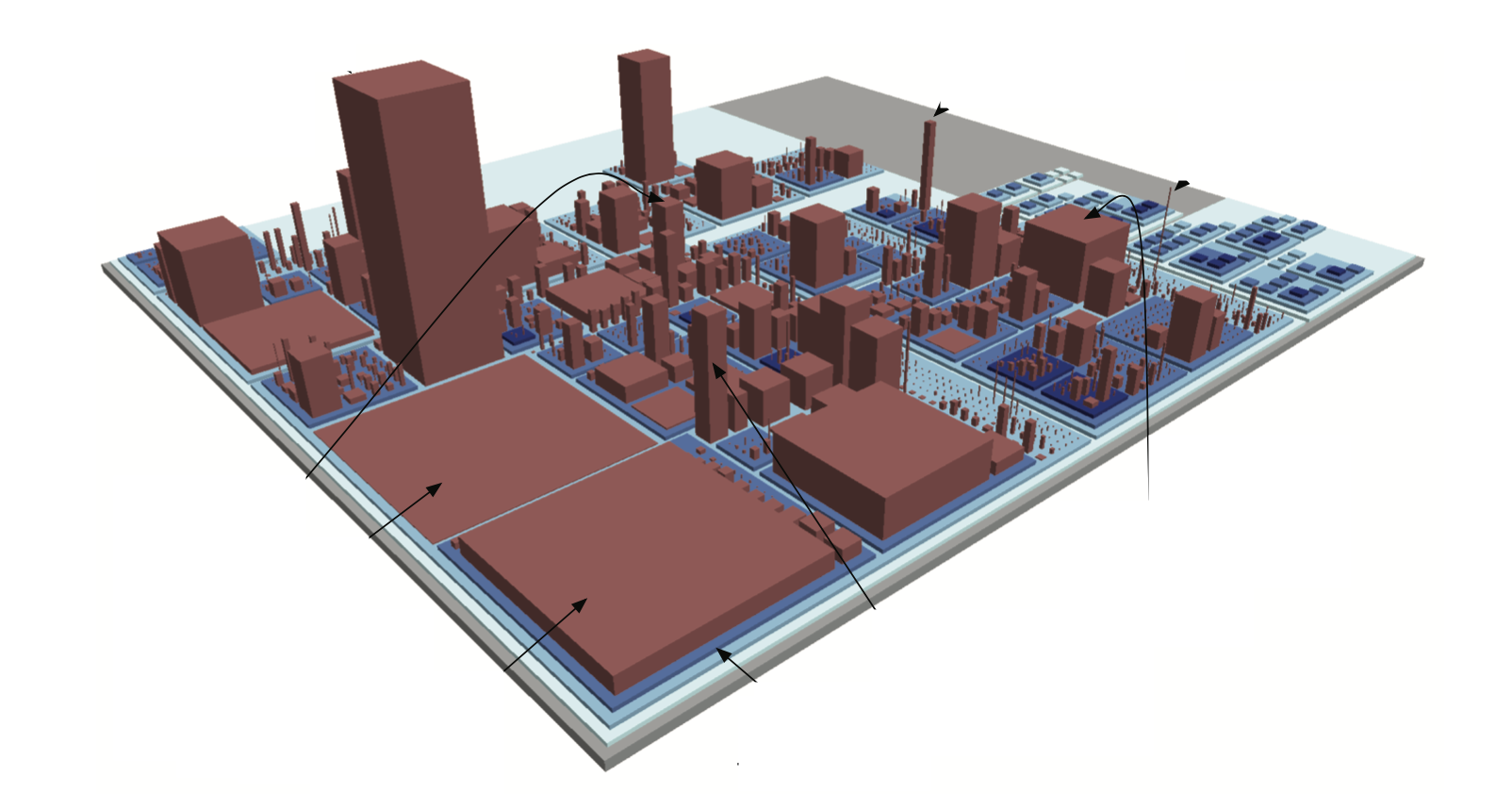}
        \caption{City Metaphore \cite{wettel2008visually}}
        \vspace{-1.5em}
        \label{fig:disharmony}
    \end{minipage}%
    \begin{minipage}{0.23\textwidth}
        \centering
        \includegraphics[width=\linewidth]{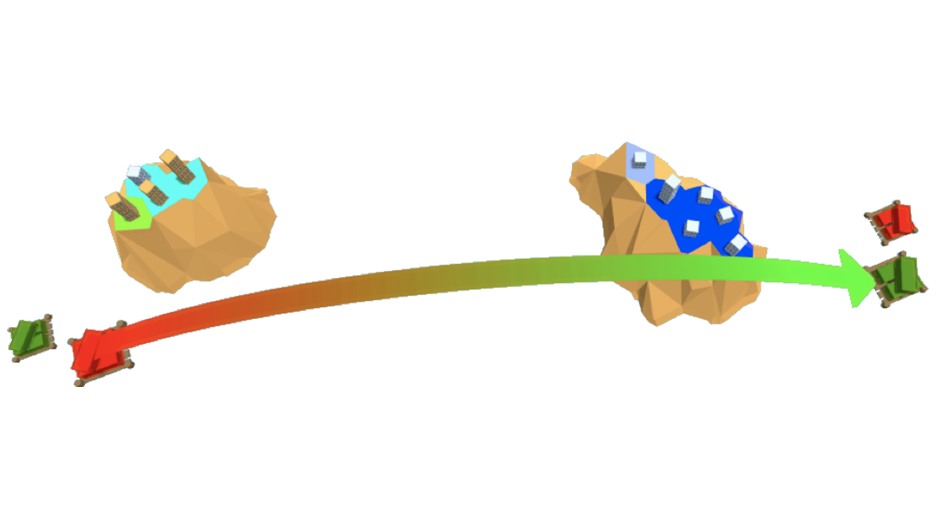}
        \caption{Island Metaphore \cite{schreiber2019visualization}}
        \vspace{-1.5em}
        \label{fig:island-metaphore}
    \end{minipage}
    \begin{minipage}{0.23\textwidth}
        \centering
        \includegraphics[width=\linewidth]{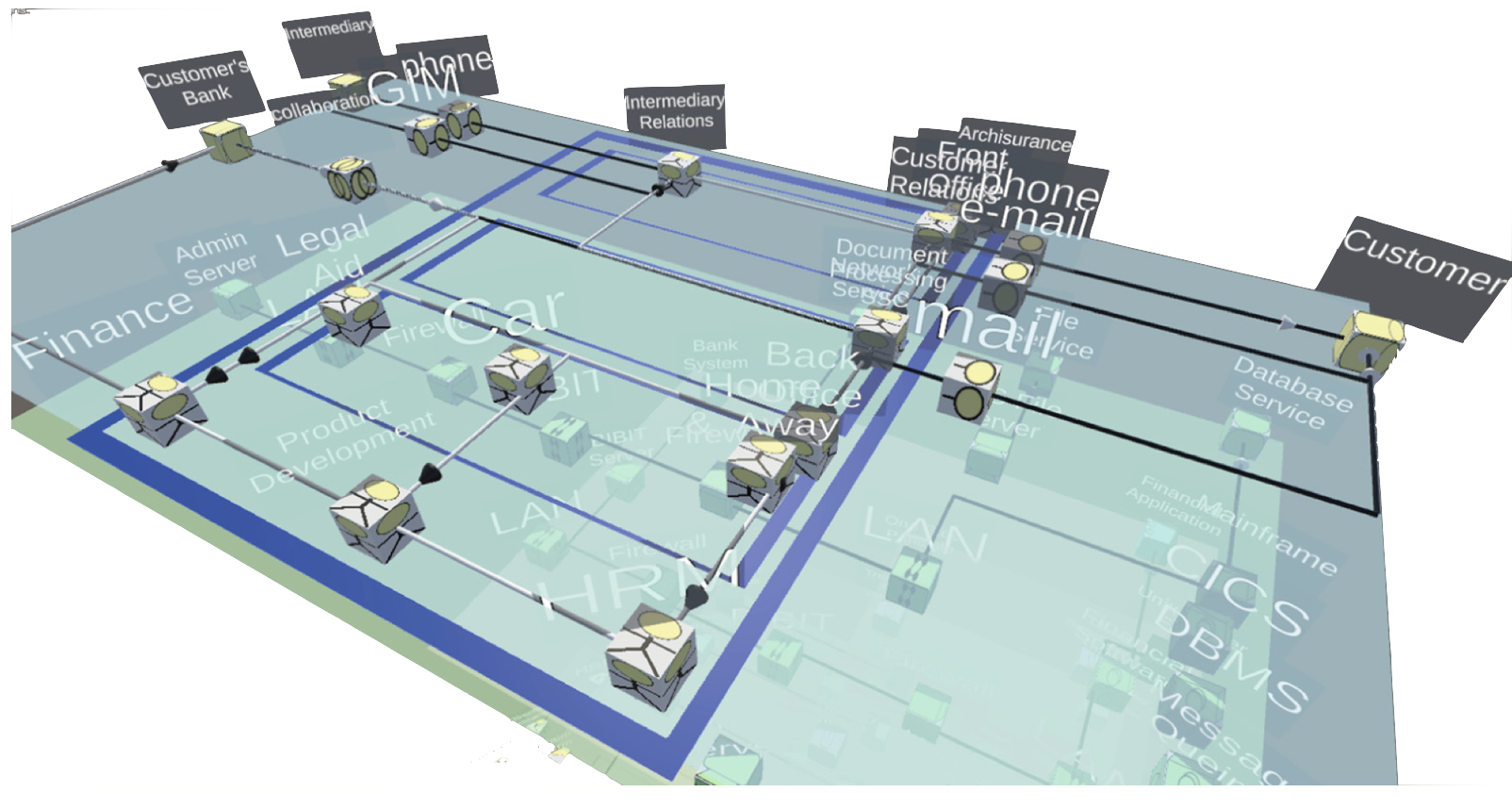}
        \caption{VR for EA Models \cite{oberhauser2019vr}}
        \vspace{-1.5em}
        \label{fig:vr-archimate}
    \end{minipage}
    \begin{minipage}{0.23\textwidth}
        \centering
        \includegraphics[width=\linewidth]{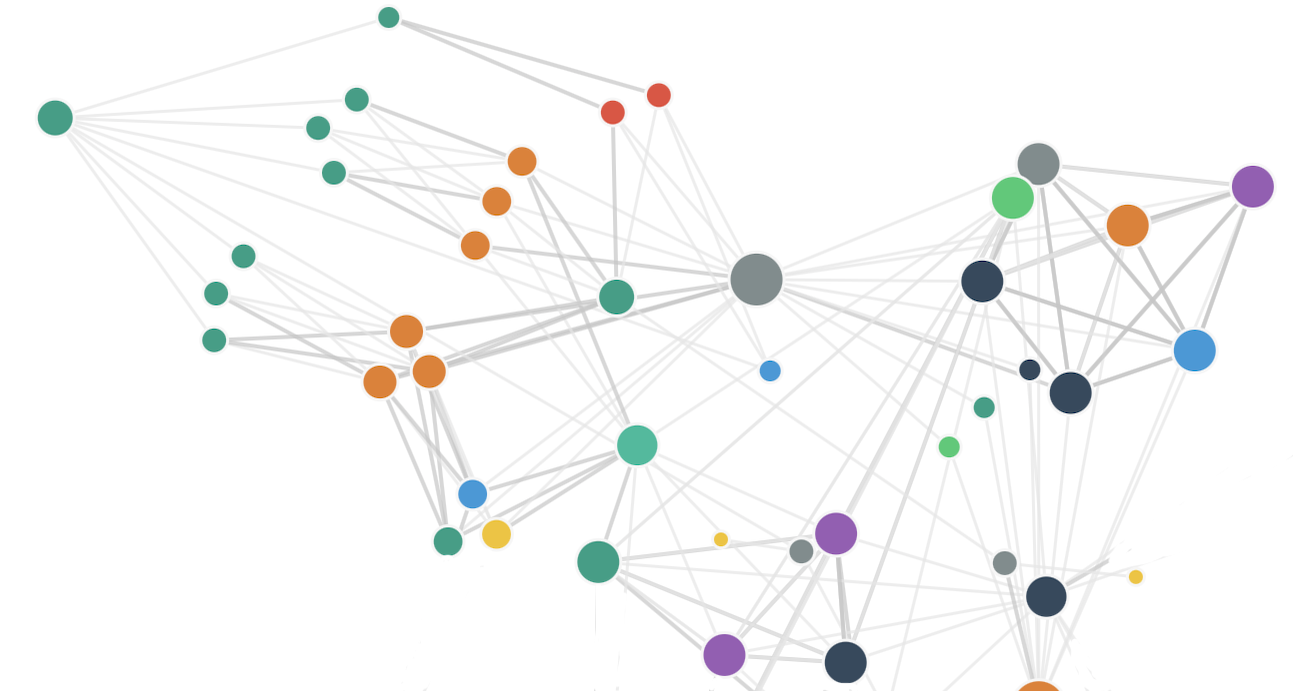}
       \caption{Netflix Demo \cite{netflixdemo}}
       \vspace{-1.5em}
       \label{fig:netflix}
    \end{minipage}
\end{figure*}
\section*{BACKGROUND AND RELATED WORK}
Multiple contributions reveal that microservices research is still in a formative stage \cite{pahl2016microservices,pease2002suggested}. Moreover, a weak research dataset did not consider crucial attributes for architecture analysis, howerver, over the past few years, research has been conducted toward proposing software visualization techniques and various taxonomies have been published. 
A lot of investigation is done in the direction of employing the notation-based UML modeling (class diagram, package diagram, interaction diagrams, and activity diagrams) for visualizing microservices-based systems, but they do not provide means for representing the concern, goal, value chain, problem, and cause \cite{zhou2020systematic}, which indicates an unscalable visualization method. Multiple contributions are invested in Enterprise Architecture (EA) Modeling for solving this scaling challenge. Zhou et al. \cite{zhou2020systematic} summarized that ArchiMate \cite{archimatespecs} is the most powerful EA modeling notation in the visual expressiveness level, as is also being the most frequently used by scholars. ArchiMate distinguishes itself from other modeling languages by its enterprise modeling scope and its ability to visualize business processes. However the ArchiMate does not consider run-time operational details of the execution as important.\\
The missing information in the modeling languages and the limitations of 2D rendering space promoted researchers to use metaphors for visualizing software architectures.
Wettel et al. \cite{wettel2008visually} presented an effective integration of the design anomaly data with a visual approach based on a 3D city metaphor as shown in Figure \ref{fig:disharmony}. It provides the big picture of the system’s design problems, however it lacks two vital features for microservices-based systems, which are traceability and dependency between services.
Metaphors play a scalable role in visualizing large systems, as shown in the island metaphor in Figure \ref{fig:island-metaphore}. Such that Schreiber et al. \cite{schreiber2019visualization} represented the entire software system as an ocean with many islands, in addition to three different levels (island, region, and building). This approach covers critical aspects in microservices architectures such as the dependency between services. However, it still requires different rendering space to digest all such details without overlapping. Therefore, the authors blended the Virtual Reality (VR) technique to overcome such limitations. The VR technique enables users to interact with 3D worlds and overcome the space limitation exists in 2D modeling.
Oberhauser et al. in \cite{oberhauser2019vr} employed the VR capability to demonstrate enterprise architecture models as shown in Figure \ref{fig:vr-archimate}. Although the VR technique helped in addressing the visualization, navigation, and interaction aspects, authors recommend finding new modeling capabilities, offering different and tailored layout options rather than the standard ArchiMate and BPMN and their many details.
Although the metaphors showed a clear view for missing aspects in UML modeling, none showed a complete solution considering navigation and traceability through different detailed levels in the system.\\
Graph representation is strongly involved as a solution. Nakazawa et al \cite{nakazawa2018visualization} proposed a graph representation visualization tool that allows developers to interactively design microservice applications. Although they targeted to balance performance and flexibility, this visualization has limitation for large-scale systems in addition to the need for missing information to be displayed.\\
Microservices still have limited scientific publication coverage in a lot of areas and grey literature may hold valuable insights that academic literature simply cannot currently provide \cite{bogner2021industry}.
Addressing the industrial contributions, Netflix shows an interactive visualization for their system \cite{netflixdemo} (Figure \ref{fig:netflix}). This service graph depicts the system topology and shows the service dependencies, at the same time it enables the user to rearrange the services to construct different topologies. Amazon provides a solution called X-Ray console \cite{xrayamazon}. It is a  visual map consisting of service nodes that serve requests, upstream client nodes that represent the origins of the requests, and downstream service nodes that represent web services and resources used by an application while processing a request. However, in the debugging scenario where a specific service is experiencing an issue, both solutions are not particularly useful.\\
While most of the existing research has focused on new visualization methods and techniques, not much work has been done to improve the usability through designing new methods. This paper addressed these challenges and recommendations to propose a novel visualization concept for microservices-based systems.
\section*{APPROACH AND UNIQUENESS}
\subsection*{Proposed Microservice Visualization Concept}
\vspace{-1.0em}
\begin{figure}[H]
\includegraphics[width=2.0in]{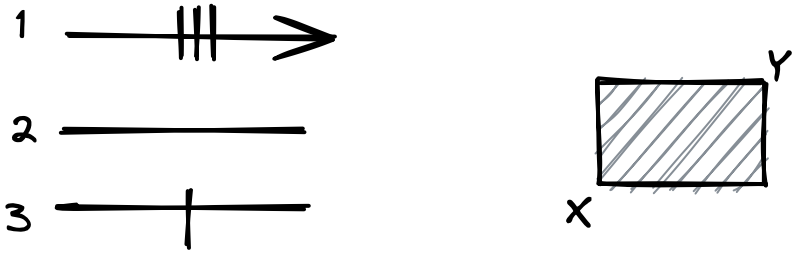}
\caption{Visualization Elements}
\label{fig:elements}
\vspace{-1.0em}
\end{figure}
\vspace{-0.2em}
\begin{figure*}[!htb]
\minipage{0.32\textwidth}
  \includegraphics[width=\linewidth]{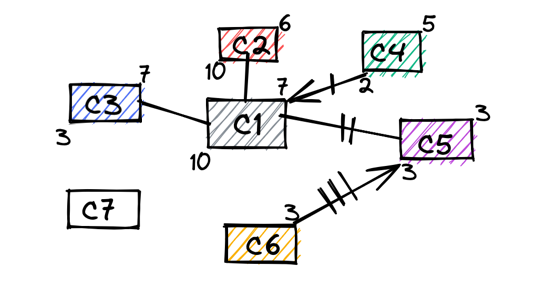}
  \caption{System Level Visualization}
  \vspace{-1.0em}
  \label{fig:highlevel-controller}
\endminipage\hfill
\minipage{0.32\textwidth}
  \includegraphics[width=\linewidth]{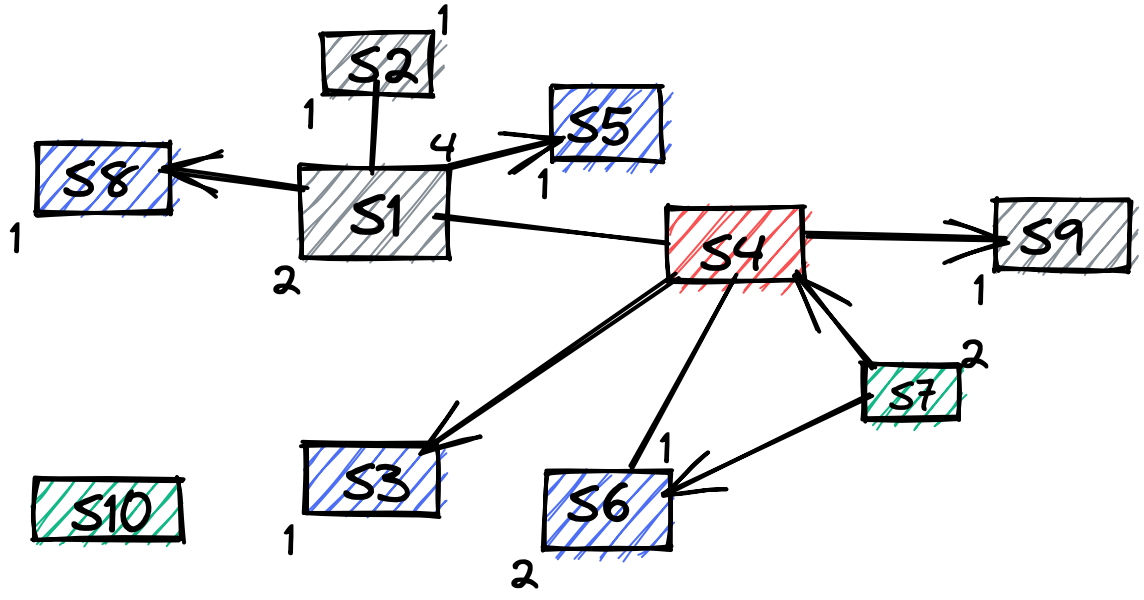}
\caption{Service Level Visualization}
\vspace{-1.0em}
\label{fig:highlevel-service}
\endminipage\hfill
\minipage{0.32\textwidth}%
\includegraphics[width=\linewidth]{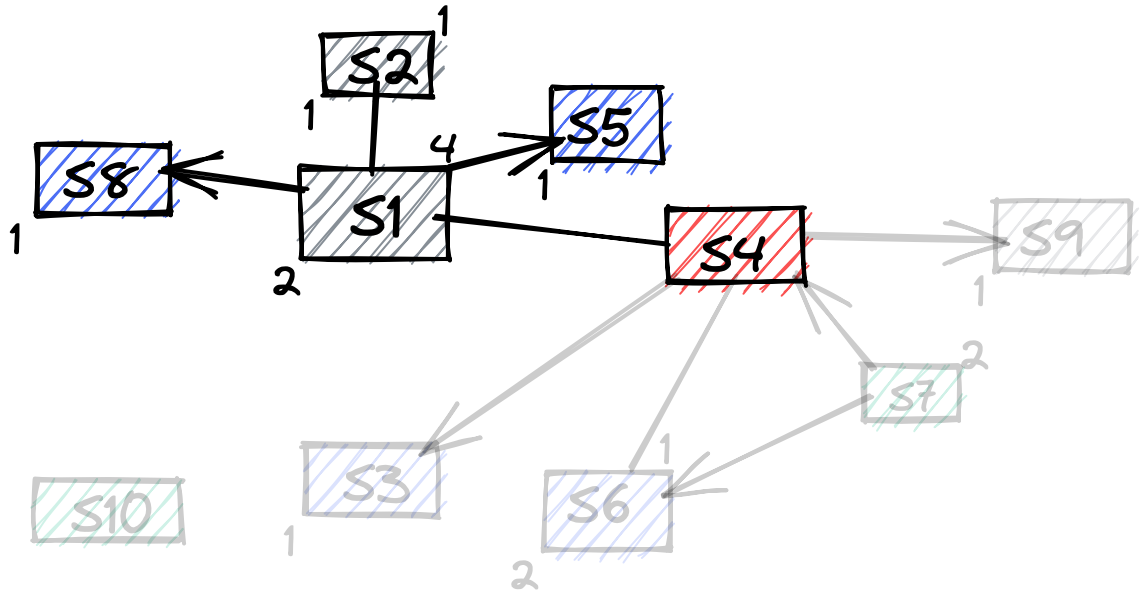}
\caption{Service Node Filter}
\vspace{-1.0em}
\label{fig:filter-node}
\endminipage
\end{figure*}
\begin{figure*}[!htb]
\minipage{0.32\textwidth}
  \includegraphics[width=\linewidth]{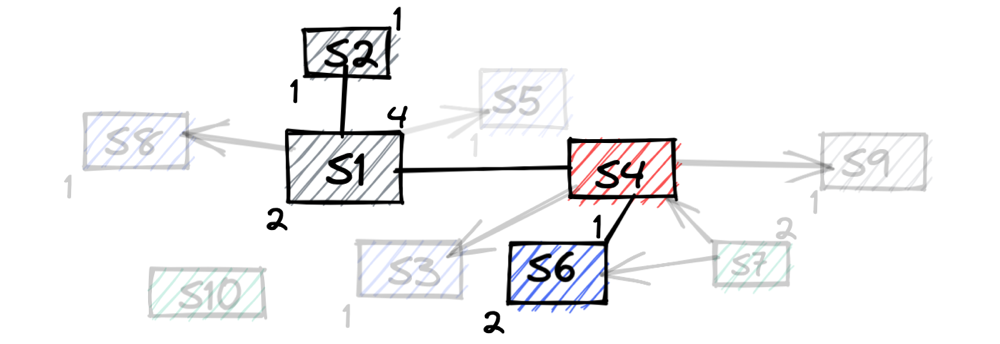}
  \caption{Path Filter}
  \vspace{-1.5em}
  \label{fig:filter-path}
\endminipage\hfill
\minipage{0.32\textwidth}
  \includegraphics[width=\linewidth]{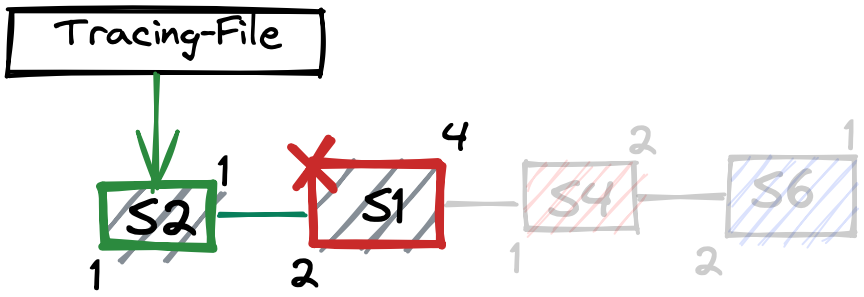}
\caption{Simulation for Service Failure}
\vspace{-1.5em}
\label{fig:simulation-path-service}
\endminipage\hfill
\minipage{0.32\textwidth}%
\includegraphics[width=\linewidth]{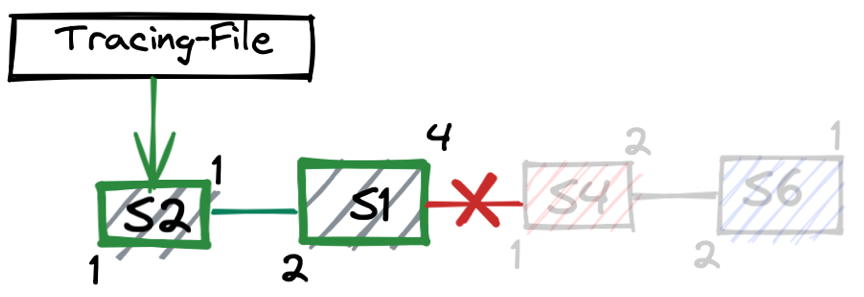}
\caption{Simulation for Path Failure}
\vspace{-1.5em}
\label{fig:simulation-path-path}
\endminipage
\end{figure*}

Visualizing and debugging complex systems, even with the increasing development of state-of-the-art tooling, is incredibly hard. Tools are required to assist a developer in the process of forming and validating a hypothesis.
This proposed concept shows user-centric, context-aware, and interactive visualization for microservices-based systems. 
The context-aware approach avoids a one-size-fits-all visualization. Therefore, this proposed concept is designed as a mobile application solution that takes the advantage of AR techniques to fit the context and overcome the rendering space limitation. This concept is also influenced by both graph representation concepts and geographical maps interactivity.\\
This section shows the illustrations of the proposed concept. Starting with the elements, as shown in Figure \ref{fig:elements}, there are two types as follows: \textit{Node} is a rectangular shape and represents the system component, which could be a service $(S)$ or a controller $(C)$. The controller is a set of services that have the same base routing. The color of the node indicates its membership, such that the same color means the same controller's members. The node size is relatively calculated using the equation of $max(X^Y, X, Y)$, such that $X$ is the number of dependant nodes to this one, and $Y$ is the number of nodes that this node depends on. This means the more the node depends on other nodes, the bigger size will be. In the event that any of these numbers are zero, the other one will be the reference. The second element is the \textit{Edge}, which indicates the dependencies between nodes. As shown in Figure \ref{fig:elements}, it is either a line or an arrow, such that edge 1 indicates unidirectional dependency, but edges 2 and 3 are for the bi-directional one. The number of small cross-lines indicates the number of participated services in this dependency. For example, if node $C1$ depends on node $C2$ on three different services, then three cross-lines will appear on their edge. For increased usability, no cross-lines will be added if there are more than three service dependencies.\\
Three different levels of architecture information are visualized through this concept, such that each level becomes increasingly detailed, as follows:
\textit{System Level} shows a high-level overview with the base of system controllers. As shown in Figure \ref{fig:highlevel-controller}, each controller has a unique color and the nodes connect with an indication about the direction as well as the number of dependencies per node. \textit{Service Level} shows the system from the perspective of services basis. As shown in Figure \ref{fig:highlevel-service}, the nodes represent services, with the same colored services belonging to the same controller that has the exact color in Figure \ref{fig:highlevel-controller}. Finally, \textit{Function Level} visualizes a communication UML diagram \cite{communication2009uml} through the AR space. It dives deeper to demonstrate the service internal functions communication flow.\\
While navigation itself is crucial, AR plays a vital role in providing it in a well-fitted render space. This concept focuses on the following navigation features:
\textit{AR Render Space}, which is the core of providing context-awareness visualization as it discovers the surrounding context with its objects as Board, Table, Wall, ... etc, and enables the user to choose which panel to use for rendering. This means that users can show the system in a different context with employing of real-life objects. \textit{Clickable Elements} enable the user to show more details through tabbing on.\textit{Draggable Elements} enable the user to rearrange the elements using drag-and-drop on-screen gestures. \textit{Zoomable Layout} is influenced by geographical maps. This allows the user to show different information corresponding to the zoom level involved. For example, nodes name will not fully appear until the user gets to certain zoom level, as will be shown below in the proof of concept, Figure \ref{fig:microvision}.\\
Moreover, this concept provides user-centric views, such that it supports multiple filtration techniques to be applied over different levels as follows: \textit{Node Filter} shows only one node with its dependencies, as $S1$ service is shown in Figure \ref{fig:filter-node}. \textit{Path Filter} highlights specific path in the view, as shown in Figure \ref{fig:filter-path} that filters the view for the path $S2 \rightarrow S1 \rightarrow S4 \rightarrow S6$. This helps the practitioner to reduce the search space.\\
All the aforementioned visualization capabilities pave the way for traceability and debugging features. This concept is designed to be a platform for microservices-based systems visualization, analysis, and traceability.
Embedding dynamic analysis information into this visualization concept proposes multiple metrics on both service and path levels. \textit{Path Hits} weighs paths according to the number of requests get through them. The user can browse them in order one by one as shown in Figure \ref{fig:filter-path}. \textit{Path Length} ranks the paths using their length. The length is calculated by the number of connected nodes per path. \textit{Service Dependency} ranks the nodes using the dependants number. This metric shows the bottlenecks from a dependency perspective, and displayed as shown in Figure \ref{fig:filter-node}.\\
Path Simulation is one of the most beneficial features of this concept. It enables users to simulate specific paths in the system for debugging and tracing purposes. Visualizing a live simulation for these paths executions to highlight failures either in the path as shown in Figure \ref{fig:simulation-path-path}, or in the service as shown in Figure \ref{fig:simulation-path-service}. This feature supports two different starting points, the first is through providing mocking data for the initial service in that path, and the second is through a tracing log file that enables automated path detection and simulation.\\
From the microservices analysis perspective, this concept provides enough slots for users to collaboratively construct different topologies. Blending AR with mobile-based technique enables different displaying contexts per user device. This gives much power to the analysis process, so that users are able to share components with each other to display these components in different context. Therefore, they could get multiple analysis scenarios and different topologies each in a separate context and a private space. \\
Multiple evaluation criteria are applied to this concept, \textit{Scalability} is supported through the three-dimensional visualization with AR space that offers better scaling with the number of services than a two-dimensional rendering space. \textit{Comprehensibility} is achieved in supporting multiple levels of detail to provide an overview of all the corresponding components. \textit{Collaboration} has been demonstrated by being a mobile-based solution that provides portability and enables sharing and interaction between users in different context. \textit{Interaction of services} shows the services behaviors in different paths. In addition to the visualized dependency details, \textit{Navigation Adherence} is supported through the deployment of device standard navigation methods, in addition to those mentioned above used to facilitate the user journey. \textit{Debugging and Traceability} are achieved through the multiple metrics and paths simulation technique for such investigation tasks. Finally, \textit{Context-aware View} is demonstrated such that the view is rendered corresponding to the current user context.
\vspace{-0.8em}
\subsection*{Proof of Concept}
\begin{figure}[h!]
\includegraphics[width=3in]{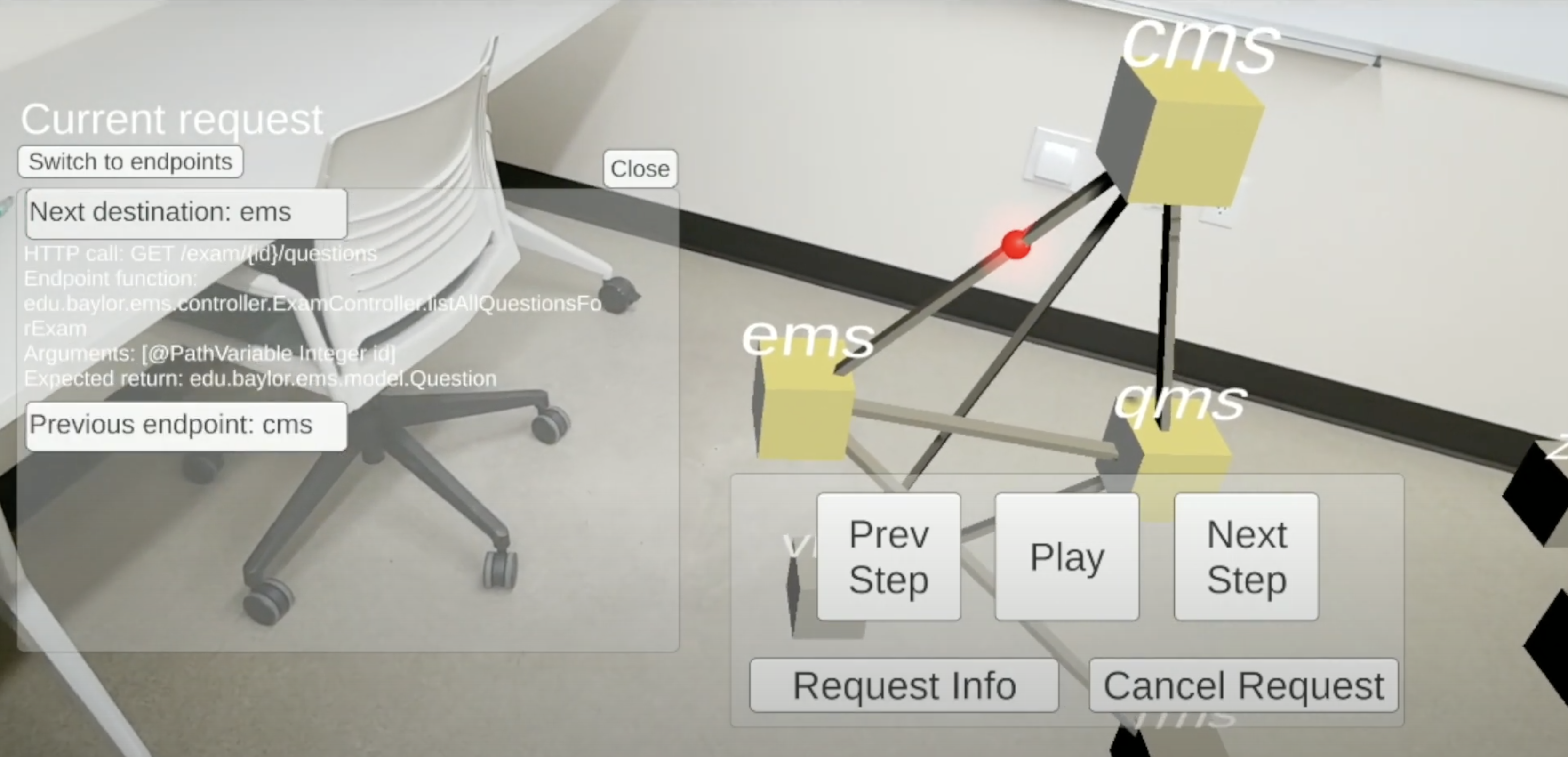}
\caption{Proof of Concept}
\label{fig:microvision}
\vspace{-2em}
\end{figure}

Our proof-of-concept is implemented as an Augmented Reality based mobile application. It addresses the aforementioned system-level view. It also supports the ability to identify details about individual services and their dependency graph in addition to primary navigation for exploring the services around the display context. As it is out of the scope of this paper, we were able to automate the Service Architecture Reconstruction (SAR)  process. We used static-code analysis to determine microservice dependencies and to analyze the microservices to construct bounded contexts of the different views. We performed this on TrainTicket testbench \cite{trainticket}, a train ticket reservation system, which contains 41 microservices. After that, the results were demonstrated through the proposed design concept as shown in Figure \ref{fig:microvision}. 
To evaluate the effectiveness of this concept in assisting developers in analyzing the architecture of a microservice system, we conducted a primary user study with six graduate student volunteers. We prepared a set of nine evaluation tasks relating to this system for the participants to complete. Three of the tasks were related to individual services and their connections in the system, and the remaining six were relating to user requests and how the system handled them. The participants used this application on an iPad Pro 11-inch device. We also prepared a five questions satisfaction survey using a 5-point Likert scale to administer to the participants regarding their experience with the application.
\vspace{-1em}
\section*{RESULTS AND CONTRIBUTIONS}
Regarding the results of the Proof of Concept study, all the participants completed the evaluation tasks with 100\% accuracy within the allotted time frame. Furthermore, all participants said the 3D graph visualization was useful, and five out of the six participants said the representation and the rendering space were useful and clear. All the participants either agreed or strongly agreed to three out of the five qualitative feedback questions. Regarding the features, the results suggest that this method is a viable platform for visualizing and analyzing microservice systems.
There were two features requests commonly reported in the final feedback question, both relating to finer-grained controls of the graph. First, participants suggested the ability to resize the graph once it had spawned. Second, they suggested the ability to move individual services within the graph. The participants indicated that these features would be helpful when trying to locate specific services. However, these features and more are already listed in the proposed concept, which emphasizes the idea that it shows viable and demanding features in an obvious way.\\
This paper contributes to the development of a novel visualization concept for microservices-based systems using AR techniques. This concept provides three different levels of views (System Level, Service Level, Function Level), two view filtration techniques (Node Filter, Path Filter), three traceability and debugging techniques (Path Hits, Path Length, Service Dependency) as well as the Path Simulation technique, and a set of navigation features in the AR rendering space. In addition to the collaboration feature between users in different contexts that has a clear vision for the analysis process.\\
This approach is still open for facing more challenges and supporting more features. The upcoming work focuses on enabling gestures interactions rather than the standard navigation functionality to increase the usability and collaboration. From the system perspective, analyzing the system for discovering and predicting architectural degradation takes advantage of this visualization to annotate and highlight these different architectural vulnerabilities. Moreover this concept has the potential to be the preferred way for producing a complete 3D modeling for microservices-based systems.
\bibliographystyle{ACM-Reference-Format}

\bibliography{sample-bibliography}
\section*{Acknowledgments}
This material is based upon work supported by the National Science Foundation under Grant No. 1854049, a grant from Red Hat Research (https://research.redhat.com)
\end{document}